\documentclass[twoside,twocolumn,tightenlines, showpacs,aps]{revtex4}
\usepackage{graphicx}
\newcommand{\be}{\begin{equation}}
\newcommand{\ee}{\end{equation}}
\begin{document}

\title{Spin Accumulation Encoded in Electronic Noise for Mesoscopic Billiards with Finite Tunneling Rates}

\author{J. G. G. S. Ramos$^1$, A. L. R. Barbosa$^2$, D. Bazeia$^3$, and M. S. Hussein$^1$}
\affiliation{$^1$ Instituto de Estudos Avan\c cados and Instituto de F\'{i}sica, Universidade de S\~{a}o Paulo,
  C.P.\ 66318, 05314-970 S\~{a}o Paulo, SP, Brazil\\$^2$ Unidade Acad\^emica de Educa\c c\~ao a Dist\^{a}cia e Tecnologia, Universidade Federal Rural de Pernambuco, 52171-900 Recife, PE, Brazil\\
$^3$Departamento de F\'\i sica, Universidade Federal da Para\'\i ba, 58051-900 Jo\~ao Pessoa, PB, Brazil}

\date{\today}

\begin{abstract}
We study the effects of spin accumulation (inside reservoirs) on electronic transport with tunneling and reflections at the gates of a quantum dot. Within the stub model, the calculation focus on the current-current correlation function for the flux of electrons injected into the quantum dot. The linear response theory used allows to obtain the noise power in the regime of
thermal crossover as a function of parameters that reveal the spin polarization at the reservoirs. The calculation is performed employing diagrammatic integration within the universal groups (ensembles of Dyson) for a non-ideal, non-equilibrium chaotic quantum dot. We show that changes in the spin distribution determines significant alteration in noise behavior at values of the tunneling rates close to zero, in the regime of strong reflection at the gates.
\end{abstract}

\maketitle
\section{Introduction}

The experimental control of electron transport in nanostructures  may lay the grounds for the development of devices for processing quantum information \cite{Qinformation1,Qinformation2,Qinformation3}. These devices may rely on the spin degrees of freedom, and are thus called spintronics \cite{Spintronics}. The control of the spin is a subtle process which requires the fabrication of special samples and manipulating them so as to detect low intensity currents in semiconductors \cite{Amostra1,Amostra2,Amostra3}. The accumulation of spin, when detected, allows the extraction of information of great value to the phenomenon of electron transport \cite{Qinformation1,Qinformation4}.

To induce a spin polarization in a material sample which can be a reservoir of electrons, one creates a population of non-equilibrium spins with a finite interval of relaxation time. This population can be achieved through optical or electronic mechanisms. Routinely, the optical techniques require the injection of circularly polarized photons in order to transfer their angular momentum to electrons through a complex sample \cite{Qinformation4}. The electronic injection involves the presence of magnetic electrodes connected to a sample, creating spin polarization in a non-equilibrium regime \cite{Qinformation3,Spintronics}.

Fluctuation properties of a non-equilibrium current indicates that just the average electronic currents are not enough for a complete description of the full quantum transport \cite{Beenakker}. The accumulation of spin in electronic reservoirs modifies the fluctuation properties of the non-equilibrium electronic current. Such a modification follow through a mechanism proposed in \cite{Jacquod} which reveals that noise power presents a asymmetry under reversal of the current/voltage in the presence of spin accumulation inside at least one reservoir. On the other hand, performing direct measurements of the fluctuations in semiconductor quantum dots can be a very hard task, precisely because the typical currents are of the order of nA and temperatures on the order of mK, very small indeed. A experimental procedure found in Ref.~[\onlinecite{gustavson1}], and  justified theoretically in [\onlinecite{levitov,nazarov}], is to perform the Full Counting Statistics (FCS) which consists of counting the numbers of electrons and their degrees of freedom within a certain window of time. Real-time measurements can also be applied to study the spin transport properties on generic interfaces of heterostructures, according to the results in \cite{gossard,Lagally}. Tunneling rates not only allow to find the conductance, but the shot-noise (width of the conductance distribution) [\onlinecite{gustavson1}].
\begin{figure}[h!]
\begin{center}
\includegraphics[width=8.0cm,height=6.5cm]{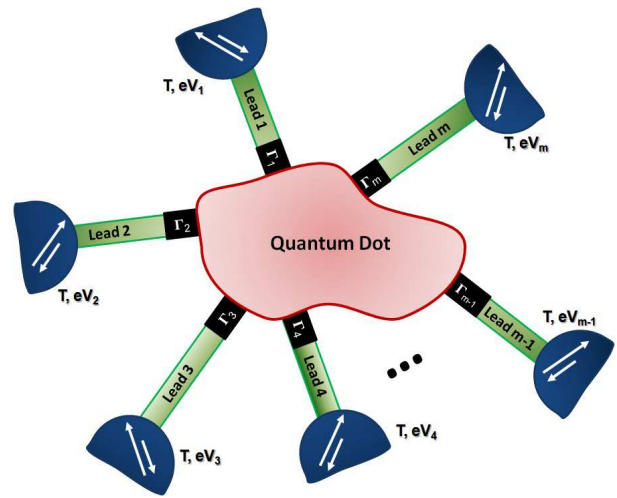}
\end{center}
\caption{A schematic picture showing a quantum dots coupled to polarizable reservoirs via several leads with open channels in the presence of temperature and voltage.} \label{ponto}
\end{figure}

In the limit of high temperatures, noise provides information on the thermal fluctuations characteristics of dissipative systems. On the other hand, experimental measurements of noise at low temperatures, also known as shot-noise, use tunneling rate in the non-ideal quantum transport [\onlinecite{gustavson1}], yielding important information about the discrete process of charge transmission [\onlinecite{blanter}]. In mesoscopic systems both noise sources are present. A relevant parameter to measure the noise in quantum dots is the asymmetry factor $a=(G_{i}-G_{j})/(G_{i}+G_{j})$, with $G_{i}\equiv N_{i}\Gamma_{i}$ and $N_{i},\;\Gamma_{i}$ denoting, respectively, the number of open channels and the tunneling rate in the lead $i$. Therefore, the tunneling rates play a crucial role in mesoscopic systems and in measures of the noise.

Motivated by these recent advances in the noise measurements [\onlinecite{gustavson1}] and by the asymmetry in current/tension seen in \cite{Jacquod}, we propose and study a myriad of possibilities to measure the spin accumulation in reservoirs through solely non-equilibrium electronic transport. This study is an alternative to that of active spin polarization in transport which usually requires the presence of ferromagnetic leads [\onlinecite{brattas}] and measurents in spin polarization through spin current is, in principle, much more difficult than measure tunneling in charge transport. For this, we consider the role of tunneling rates in the electronic transport for quantum dots coupled to reservoirs through normal guides. Considering independent electron spin distributions of these reservoirs, we show that the average noise displays new and surprising effects due to the asymmetry parameter $a$. There are many theories for the calculation of electron counting statistics. To name a few: Non-linear $\sigma$ models (replica \cite{m1}, supersymmetric \cite{m2} and Keldysh \cite{m3}), quantum circuit theory {\cite{m4}}, cascade approach \cite{m5}, stochastic path integral technique\cite{m6}, semi-classical methods based on solving Boltzmann-Langevin equations\cite{m7}, etc. In this paper, we use one more, and proven powerful, method based on the Random Matrix Theory (RMT). More specifically, using RMT \cite{mehta,Beenakker} we study the generalization of the interesting experimental setup recently proposed in Ref.~[\onlinecite{Jacquod}].

We consider an open QD connected to $m$ reservoirs labeled by $\alpha=1,\dots,m$ through leads with open electronic channels. The system, schematically represented in Fig. 1, contains reservoirs with electro-chemical potentials $\mu_{\alpha}=\mu_{\alpha \uparrow}+\mu_{\alpha \downarrow}$, where $\uparrow$ and $\downarrow$ denoting, respectively, the contributions of spins up and down. The reservoirs are kept at an arbitrary temperature $ T $ in a way that the system can reach the thermal crossover. The tunneling rates, $\Gamma_{i}$, can be controlled through changes in the gates voltage. We consider non-equilibrium corrections at the electro-chemical potentials owing to the accumulation of spin. This is denoted by $\delta \mu_{\alpha}=(\mu_{\alpha \uparrow}-\mu_{\alpha \downarrow})/2$. Because of the difference $n_{\alpha \uparrow} - n_{\alpha \downarrow}$ at the reservoirs, where $n_{\alpha}$ is the total number of electrons, there is a well defined direction of the spin polarization at, say, reservoir $\alpha$, which we describe by the unit vector ${\bf m}_{\alpha}$. We show that tunneling rates drastically affects the measurements in spin distribution at reservoirs of QD. We further show that a great change in the average noise power occurs in a region of spin accumulation close to where most experiments have performed.

\section{Scattering Theory of Quantum Transport}\label{sistemaquantico}

In Section A, we will make a brief presentation of the linear response theory using Landauer-B\"uttiker scattering formalism. We follow \cite{Jacquod} that verify the asymmetry current/voltage and present their main results, making our work self-contained. The theory presented includes an arbitrary topology and the separation of spin degrees of freedom. In section B, we present original results for average noise including the spin accumulation in the presence of tunneling rates.

\subsection{General Formulation}

In the limit of low bias voltages, we construct a theory of multi-terminal and multi-channel scattering, generating the Landauer-B\"{u}uttiker framework for quantum transport \cite{blanter}. We start by considering the time-dependent current $\hat{I}_{\gamma}(t)$ at lead $\gamma$, for $\gamma=1,2,...,m$, with $m$ being the number of leads connected to the chaotic quantum dots. Within the framework of the scattering theory for quantum transport, the current-current correlation function can be written in the form \cite{blanter}
\begin{eqnarray}
\langle \delta \hat{I}_\alpha(t) \delta \hat{I}_\beta(0)\rangle=\int\frac{dw}{2\pi}e^{-iwt}\mathcal{S}_{\alpha \beta}(w), \label{fourier}
\end{eqnarray}
where $ \delta \hat{I}_\alpha(t)\equiv  \hat{I}_\alpha(t)+\langle \hat{I}_\alpha(t)\rangle$ is the current fluctuation around the mean value $ \langle \hat{I}_\alpha(t)\rangle$. The Fourier transform of the current-current correlation function, Eq.(\ref{fourier}), namely $\mathcal{S}_{\alpha \beta}(w)$, is the noise, which, in the absence of interaction, can be written as \cite{Jacquod}
\begin{eqnarray}
\mathcal{S}_{\alpha\beta}(w)&=&\sum_{\gamma \nu} \sum_{(c_{1},p)\in \gamma} \sum_{(c_{2},q)\in \nu} \frac{e^2}{h}\int d\varepsilon\;  A_{\gamma \nu}^{c_{1},p;c_{2},q}(\alpha;\varepsilon,\varepsilon')\nonumber\\&&\times A_{\nu\gamma}^{c_{2},q;c_{1},p}(\beta;\varepsilon', \varepsilon)\left\{f_\gamma^{p}(\varepsilon)\left[1-f_\nu^{q}(\varepsilon')\right]\right. \nonumber
\\&&\left.+f_\nu^{q}(\varepsilon)\left[1-f_\gamma^{p}(\varepsilon')\right]\right\}; \;\; \varepsilon' \equiv \varepsilon+\hbar w.\label{ruido}
\end{eqnarray}
The matrix $A_{\gamma \nu}^{c_{1},p;c_{2},q}(\alpha;\varepsilon, \varepsilon') \equiv \delta_{c_{1}c_{2}}\delta_{pq}\delta_{\alpha \gamma}\delta_{\alpha \nu}-[S^{\dagger}_{\alpha \gamma}(\varepsilon)S_{\alpha \nu}(\epsilon')]_{c_{1},p;c_{2},q}$ is the current matrix, where $S(\varepsilon)$ is the scattering matrix, which can depend on the energy $\varepsilon$ and describes the charge transport through the circuit. Also, $f_\gamma^p(\varepsilon) = \left (1+ \exp \left [(\varepsilon- \mu_{\gamma p}) / k_BT \right] \right)^{- 1}$ represents the Fermi distribution function, related to the thermal reservoir connected to the lead $\alpha$. The sum in Eq.(\ref{ruido}) extends over spin indices $ p, q = \pm $ polarizable along $ {\bf m} _ {\gamma} $, open channels indices $ c_{1}, c_{2} \in \gamma $ and over all leads, including $ \alpha $ and $ \beta $.

The scattering matrix $S(\varepsilon)$ used to describe the mesoscopic system is uniformly distributed over the orthogonal ensemble, if the system has both time-reversal and spin rotation symmetry, over the unitary ensemble, if only time-reversal symmetry is broken by a intense external magnetic field, or over the symplectic ensemble, if the spin rotation symmetry is broken by a intense spin-orbit interaction \cite{mello}.

A particularly interesting limit of the resulting linear response theory is that at zero frequency, for which there is a successful model established to treat noise of a phase-coherent conductor \cite{buttiker}. In this limit, we define $\mathcal{S}_{\alpha\beta}=\mathcal{S}_{\alpha\beta}(0)$ and the transport is described in terms of external fields contained in the symmetries of the scattering matrices, the energies present in the corresponding Fermi distributions in the reservoirs, and on the open channels in the leads. In the limit of both low temperatures and voltages, the scattering matrix is uniform within an energy windows in the vicinity of Fermi level, in a form that the scattering matrix is given by $S=S(\varepsilon)=S(E_{F}), \; \forall \varepsilon$, with $E_{F}$ denoting the Fermi energy. From Ref.~[\onlinecite{buttiker}], along with the limits discussed above, spectral noise of the current-current correlation function function can be written as \cite{Jacquod,dragominova}
\begin{eqnarray}
\mathcal{S}_{\alpha\beta}&=& 2k_B T\left[\delta_{\alpha\beta} 2 N_\alpha-\text{Tr}\left(1_\beta S^{\dagger}1_\alpha S+1_\alpha S^{\dagger}1_\beta S\right)\right]\label{noise}\\
&+&\frac{1}{4}\sum_{\gamma,\rho=1}^m \sum_{p,q=\pm} {\it f}_{\gamma\rho}^{pq}\left[ \mathcal{T}_{\gamma\alpha\rho\beta}^{00}+2 p \text{Re}
\mathcal{T}_{\gamma\alpha\rho\beta}^{z0}+p q\mathcal{T}_{\gamma\alpha\rho\beta}^{zz}\right]; \nonumber \\
f_{\gamma\rho}^{pq}&\equiv&\int dE \left[f_\gamma^p\left(1-f_\rho^q\right)+f_\rho^q\left(1-f_\gamma^p\right)\right].\nonumber
\end{eqnarray}
The matrix $S$ has dimension $ 2M \times 2M$, with $M=\sum^m_\gamma N_\gamma$  denoting the total number of open channels in the leads. The matrix $1_\alpha $ projects states on the transport guide $ \alpha $. We also define
\begin{eqnarray}
 \mathcal {T} _ {\gamma \alpha \rho \beta} ^ {ab} \equiv \text {Tr} \left [\left (1_ \gamma \otimes \sigma^a \right) S^ {\dagger} 1_\alpha S \left (1_\rho \otimes
\sigma ^b \right) S^{\dagger} 1_\beta S \right] \label{spinparte}
\end{eqnarray}
where $a,b \in \{ 0,z \}$, $ \sigma^z = {\bf \sigma} \cdot {\bf m} _ \rho $ with $ {\bf \sigma} $ is the Pauli vector/matrix and $\sigma^0$ a identity matrix $2 \times 2 $.

\subsection{Non-ideal Mesoscopic Billiards}

Now, we present our new results, extending \cite{Jacquod} to include tunneling and reflections. The scattering matrix incorporates the non-ideal coupling between the ideal-channels of the leads and the internal modes of the QD. This coupling describes the  tunneling rate $ \Gamma_{\alpha} \in [0,1] $ of the entrance and exit of the electronic modes of lead $ \alpha $ in the QD. In RMT, the tunnel rate is generically referred to as tunneling barrier. The presence of barriers imposes a distribution of the scattering matrices within the Poisson kernel \cite{mello,Beenakker,brouwer,datta} of RMT and integration in the Haar measure corresponding to extracting non-analytical results for the averages. Therefore we will use the diagrammatic method proposed in Ref.~[\onlinecite{brouwer}], to find the leading term in the semi-classical expansion of the average noise. Following Refs.~[\onlinecite{nos,brouwer}], the matrix $ S $ can be parameterized by the stub model, being composed by an average part, $R$, and a fluctuating part, $\delta S$:
\begin{eqnarray}
S&=&R+\delta S \nonumber\\
\delta S&=&T[1-RU]^{-1}UT^\dagger.\nonumber
\end{eqnarray}
The matrix $U$ is random orthogonal, unitary or sympletic, depending on the Dyson ensemble, with dimensions $2M\times 2M$. The matrices $T$ and $R$ are diagonal, $2M\times 2M$, matrices, given by $T=\text{diag}\left(i\sqrt{\Gamma_1}1_{2N_1}, \ldots, i\sqrt{\Gamma_m}1_{2N_m}\right)$ and $R=\text{diag}\left(i\sqrt{1-\Gamma_1}1_{2N_1}, \ldots, i\sqrt{1-\Gamma_m}1_{2N_m}\right)$.

In the limit of many open channels $ M \gg 1 $, we can expand $ S $ in powers of $ U $ and perform a diagrammatic integration, obtaining average moments of the scattering matrix in the Poisson kernel. According with Eq. (\ref{noise}), the average noise requires the calculation of the semiclassical expansion of the trace of products of two and four scattering matrices. We performed the calculation and verified explicitly that only the ladder diagrams (difusons) contribute to the leading term of the average noise. The diagrams for the average over trace of the product of two $ S $ matrices can be found in Ref.~[\onlinecite{brouwer}], while the diagrams for obtaining the average of four matrices $ S $ can be found in Ref.~[\onlinecite{nos}]. We get for a ballistic chaotic quantum dot connected to multiple terminals the following known general result:
\begin{eqnarray}
\langle\text{Tr}\left(1_\beta S^{\dagger}1_\alpha S\right)\rangle=2\delta_{\beta\alpha} \left[N_\beta-G_\beta\right]+2\frac{G_\beta G_\alpha}{G_T}.\label{media2S}
\end{eqnarray}
The average of Eq.(\ref{spinparte}) is calculated in a generic form for any ensemble, arbitrary number of leads and different tunneling rates in each lead. We obtain the following, new, result valid for the universal ensembles:
\begin{widetext}
\begin{eqnarray}
\langle\mathcal{T}_{\gamma\alpha\rho\beta}^{ab}\rangle &=&2\delta_{ab}\left\{\delta_{\gamma \alpha \rho \beta}  \left(N_\gamma-2 G_\gamma+G_\gamma \Gamma_\gamma\right)+\delta_{a0}\frac{G_\gamma G_\alpha G_\rho G_\beta }{G_T^3} \left[2-\Gamma_\gamma-\Gamma_\alpha-\Gamma_\rho-\Gamma_\beta+\frac{\sum_{i=1}^m G_i\Gamma_i  }{G_T}\right]\right.  \label{media4S}\\
&+&\left.\frac{G_\gamma G_\alpha }{G_T}\left[\delta_{\gamma \rho \beta}\left(1-\Gamma_\gamma\right)+\delta_{a 0}\delta_{\alpha \rho \beta}\left(1-\Gamma_\alpha\right)\right]+\frac{G_\rho G_\beta}{G}\left[\delta_{\gamma \alpha\rho }\left(1-\Gamma_\rho\right)+\delta_{a 0}\delta_{\gamma\alpha\beta}\left(1-\Gamma_\beta\right)\right]\right.\nonumber\\
&+&\left.\frac{G_\gamma G_\alpha G_\beta }{G_T^2}\left[\delta_{\gamma\rho}\Gamma_\gamma-\delta_{a 0}\frac{G_\rho }{G_\beta}\left[\delta_{\gamma \beta}\left(1-\Gamma_\gamma\right)+\delta_{\rho \beta}\left(1-\Gamma_\rho\right)-\delta_{\alpha\beta}\Gamma_\alpha\right]-\delta_{a0}\frac{ G_\rho}{G_\alpha}\left[\delta_{\gamma \alpha }\left(1-\Gamma_\gamma\right)+\delta_{\alpha\rho}\left(1-\Gamma_\rho\right)\right]\right] \right\},\nonumber
\end{eqnarray}
\end{widetext}
where $G_m=N_m\Gamma_m$, $G_T=\sum_{i=1}^m G_m$ and $\gamma, \alpha,\rho,\beta=1,\dots,m$. In unitary or symplectic ensembles, we consider the non-colinear spin accumulation in the direction of the unit vectors ${\bf m}_{\gamma}$ e ${\bf m}_{\rho}$ such that we replace ${\bf m}_\gamma\cdot{\bf m}_\rho \rightarrow \delta_{ab}$, where $a = z= b$. In the orthogonal case, we should take $\delta_{a0} \rightarrow 1$ owing to both spin rotation and time-reversal symmetries. In the case of absence of spin accumulation, for which the equations (\ref{media2S}) and (\ref{media4S}) can be used with $a=0=b$, we recover the known results of literature [\onlinecite{epl}]. We also recover the ideal contacts case, $\Gamma_{i}=1$, in the presence of spin accumulation obtained in [\onlinecite{Jacquod}] for the average noise. Our general result is the main (semiclassical) term of the average noise and it is valid for three ensembles of Dyson. Without loss of generality, we focus on unitary ensemble and study surprising asymmetries due to tunneling rates. Sample-to-sample measurements can lead to corrections discussed in ref. [\onlinecite{Jacquod}], which gives rise to another noise asymmetries, from the $\mathcal{T}_{\gamma\alpha\rho\beta}^{0z}$ term in Eq. (3) (zero in average).
\\
\begin{center}
\begin{figure}
\includegraphics[width=9.0cm,height=7.0cm]{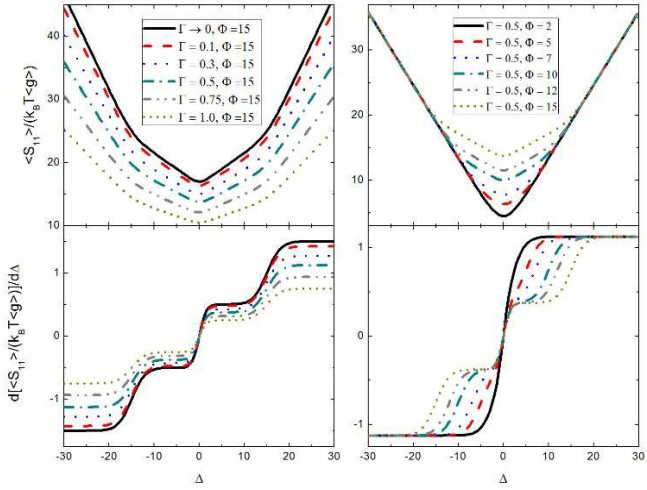}
\caption{We depict the behavior of the noise and its first derivative in the regime of spin accumulation as a function of tunneling rate $\Gamma$ in a quantum dot with non-ideal symmetrical contacts. For $\Delta > \Phi$,  We observe that the noise suffers abrupt changes, enhanced by the finite tunneling rates.} \label{ruidobarreirassimetricas}
\end{figure}
\end{center}

\section{Electronic noise power and spin accumulation in reservoirs}

The previous results are general and apply to the case of many terminals coupled to the QD. In this section, we  analyze the more widely studied case of noise in the regime of spin accumulation in a ballistic QD coupled to two leads with non-ideal contacts as the described in figure \ref{ponto}. We study in details the two terminals case, having in mind the curious and surprising fact that this configuration presents a clear instance of non-equilibrium spin accumulation phenomena, quite interesting for direct phenomenology and investigations of most usual experiments on noise. Let us consider a number of open channels $ N_ {1} $ and $ N_ {2} $ in the leads are connected to the reservoirs labeled by $ 1 $ and $2 $, respectively. Without loss of generality, we assume that the accumulation of spin occurs only in reservoir $ 1 $ so that $ \Delta \mu_1 = eV + p \delta \mu $ and $ \delta \mu_2 = $ 0 in the Fermi distribution. Substituting the general equations (\ref {media2S}) and (\ref{media4S}) in equation (\ref{noise}), we get the expression  for the two terminals case:

\begin{widetext}
\begin{eqnarray}
\frac{\langle\mathcal{S}_{11}\rangle}{k_BT\langle g\rangle}&=&\frac{6G_1G_2}{G_T^2}
+\frac{G_1G_2\Gamma_1\left(2G_2+G_1\right)}{G_T^3}+\frac{4G_2^3\Gamma_1+3G_1^3\Gamma_2}{G_T^3}\nonumber\\
&+&|\Delta|\coth\left(|\Delta|\right)\left[\frac{G_1^2G_2\left(2-\Gamma_1\right)}{G_T^3}
+\frac{2G_1G_2^2\left(1-\Gamma_1\right)}{G_T^3}+\frac{G_1^3\Gamma_2}{G_T^3}\right]\nonumber\\
&+&\left[|\Phi+\Delta|\coth\left(\frac{|\Phi+\Delta|}{2}\right)+|\Phi-\Delta|\coth\left(\frac{|\Phi-\Delta|}{2}\right)\right]
\left[\frac{G_1G_2}{G_T^2}+\frac{G_1^3\left(1-\Gamma_2\right)+G_2^3\left(1-\Gamma_1\right)}{G_T^3}\right]\label{ruidospin}
\end{eqnarray}
\end{widetext}
with $\langle g\rangle=2G_1G_2/G_T$, $\Phi=eV/k_BT$ and $\Delta=\delta\mu/k_BT$. We also show that this equation satisfies the conservation law $S_{i1}=-S_{i2}$, with $i=1,2$, indicating that the behavior of any $S_{ij}$ is identical.

Before we analyze equation (\ref{ruidospin}), we should first verify  several of its basic limits. We start by considering the limit $k_BT \gg eV, \delta \mu$ and obtain the universal thermal noise $\langle\mathcal{S}_{11}\rangle=4k_BT\langle g\rangle$. Another important case which leads to the shot-noise power is the limit  $eV \gg \delta \mu, k_BT$, through which we find that $F=\langle\mathcal{S}_{11}\rangle/2eI$ where $F$ is the Fano factor and $2eI$ is the Poisson noise:
\begin{eqnarray}
F=\frac{G_1G_2}{G_T^2}+\frac{G_1^3\left(1-\Gamma_2\right)+G_2^3\left(1-\Gamma_1\right)}{G_T^3}. \label{fano}
\end{eqnarray}

From Eq.(\ref{fano}), we can see that the case of symmetric contacts, $G = G_1 = G_2$ and $\Gamma_1 = \Gamma_2 = \Gamma $, the Fano factor simplifies to $ F = 1 / 4 \times (2 - \Gamma) $, under typical ballistic QD for which $ F = 1 / 4 $ in the case of ideal contacts. It is also possible to see in Eq. (\ref{ruidospin}) that the noise is non-zero even when $eV \rightarrow 0$ for an arbitrary value of temperature crossover. The spin accumulation maintains the noise for arbitrary electrochemical potentials for both shot-noise power and thermal noise power. The general Eq.(\ref{ruidospin}),  in the case of symmetric contacts simplifies to the following expression:
\begin{widetext}
\begin{eqnarray}
\frac{\langle\mathcal{S}_{11}\rangle}{k_BT\langle g\rangle}=\frac{6+5\Gamma}{4}
+\frac{2-\Gamma}{4}\left[|\Delta|\coth\left(|\Delta|\right)+|\Phi+\Delta|\coth\left(\frac{|\Phi+\Delta|}{2}\right)+|\Phi-\Delta|\coth\left(\frac{|\Phi-\Delta|}{2}\right)\right]. \label{ruidosimetrico}
\end{eqnarray}
\end{widetext}

The behavior of equation (\ref{ruidosimetrico}) is displayed in Figure (\ref{ruidobarreirassimetricas}). In the left figure, we fix $\Phi$ at a fixed generic value and also fix several values of the barriers. We observe in this figure that the barrier greatly amplify the signal of $ \left< S_{11} \right>/k_{B}T \left< g \right>$. We observe two anomalous characteristics of the first derivative: The first centered at the inversion point of the spin polarization of the reservoir, and the second in the region of saturation at which $\Phi=\Delta$. In these zones drastic changes of the rate of increase in the noise, encoded in the value of its first derivative which stabilizes between two plateaus as the bias voltage decreases. In the right figure, we investigate the finite value of $\Gamma = 0.5$ of the tunneling rate and the disappearance of one of the plateaus. The elimination of one of the plateaus of the first derivative indicates that the tunneling rate has an important role in the study of the saturation zone as the bias voltage is decreased. It is one of the important effects of the tunneling rate on the spin accumulation in the system. Taking the limit $ \delta \mu \gg eV, k_BT $, we obtain
\begin{eqnarray}
\frac{\langle\mathcal{S}_{11}\rangle}{\langle g\rangle}=\frac{3}{4}\left(2-\Gamma\right) |\delta\mu|,
\end{eqnarray}
which can be rewritten in terms of the Fano factor as $\langle\mathcal{S}_{11}\rangle/\langle g\rangle=3\times F \times |\delta\mu|$.
\begin{center}
\begin{figure}
\includegraphics[width=8.6cm,height=8.0cm]{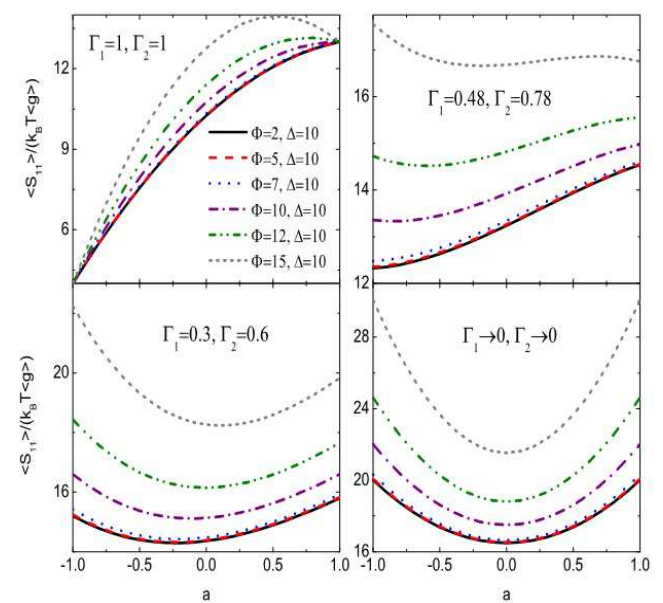}
\caption{We depict the behavior of the noise in the regime of spin accumulation as a function of the tunneling rate $\Gamma$ in a QD with non-ideal contacts, through the parameter  $a=\left(G_1-G_2\right)/G_T$. Note that for ideal contacts, $\Gamma_1=\Gamma_2=1$, the noise is highly asymmetrical with respect to this parameter. In the other curves we vary the values of $\Gamma_1$ and $\Gamma_2$  till we reach the opaque limit, $\Gamma_{1,2} \rightarrow 0 $. In this case the noise becomes symmetrical with respect to $a$. } \label{ruidobarreirasassimetricas}
\end{figure}
\end{center}

\section{Opaque Limit}

A particularly interesting regime in experiments involving tunneling rates is called ``Opaque Limit". The experimental data in real-time traces of Refs.~[\onlinecite{gustavson1,gossard,Lagally}] are basically in this category. The opaque limit is well-defined in Ref.~[\onlinecite{Whitney}], where analytical calculation using semiclassical method were performed allowing the obtention of time scales typical of transport phenomena in ballistic cavities. This regime is defined by taking limits of $ N_\alpha \rightarrow \infty $ and $ \Gamma_ \alpha \rightarrow 0 $ such that $ G_\alpha $ be finite. Taking this limit, the general expression (\ref{ruidospin}) simplifies to the following equation
\begin{eqnarray}
\frac{\langle\mathcal{S}_{11}\rangle}{k_BT\langle g\rangle}&=&\frac{\left(1-a^2\right)}{2}\left[3
+|\Delta|\coth\left(|\Delta|\right)\right] \nonumber \\
&+&\frac{\left(1+a^2\right)}{2}\left[|\Phi+\Delta|\coth\left(\frac{|\Phi+\Delta|}{2}\right)\right.\nonumber\\
&+&\left.|\Phi-\Delta|\coth\left(\frac{|\Phi-\Delta|}{2}\right)\right], \label{limiteopaco}
\end{eqnarray}
where we have defined  $a \equiv \left(G_1-G_2\right)/G_T$, thus totally encoding the open channels. The entrance and exit events of the QD are uncorrelated and the asymmetric parameter $a$ of the tunneling rate was used in Ref.~[\onlinecite{gustavson1}] to designate the normalized moments of a single level in the QD.

\begin{figure}
\begin{center}
\includegraphics[width=7.0cm,height=7.0cm]{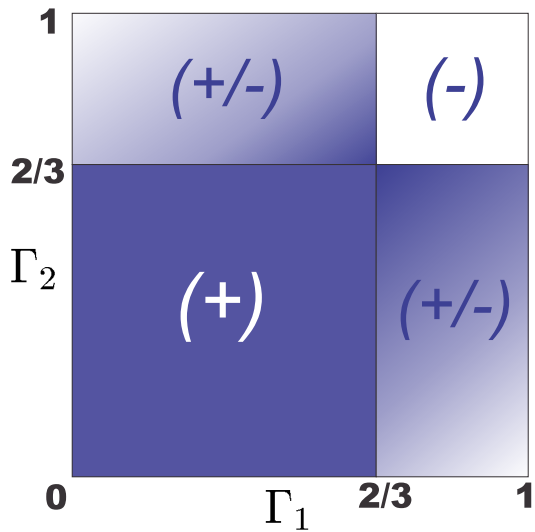}
\end{center}
\caption{In this figure we show the concavity and the sign of the noise power for all values of the tunneling rate and of the asymmetry parameter.} \label{second2}
\end{figure}

In Figure (\ref{ruidobarreirasassimetricas}), we analyze how the tunneling rates affects the noise, equation (\ref{ruidospin}), through the $ a $ parameter. Note that for ideal contacts, $ \Gamma_1 = \Gamma_2= 1 $ the noise is highly asymmetrical with respect to this parameter. This asymmetry is a result of of the spin accumulation in QD, considering that the noise is always symmetrical with respect to $a$ in the absence of spin accumulation, regardless to the values of $\Gamma_1$ and $\Gamma_2$. A similar asymmetry effect owing to the topology of the QD was reported in [\onlinecite{prbnew}]. In the other curves shown in this figure (\ref{ruidobarreirasassimetricas}), we present the behavior of the noise when we vary the values of the parameters $\Gamma_1$ and $\Gamma_2$ till we reach the opaque limit, $\Gamma_ {1,2} \rightarrow 0 $. In the transition  between the two regimes, we found that decreasing the tunneling rates has the effect of symmetrizing the noise. Surprisingly, in the opaque limit the noise becomes symmetric with respect to $a$ even in the presence of spin accumulation at any value of $\Delta$. Namely, the opaque limit symmetrizes the noise with spin accumulation, and the control parameter  responsible for this transition is the tunneling rate exemplified by $\Gamma_i$. In addition, once again we find that for values such that $ \Delta> \Phi $, the noise remained stationary in terms of $a$.

We note here that the ideal-opaque transition, determined by the finite value of the tunneling rate, also  inverts the concavity of the noise signal. The second derivative of the noise as a function of the asymmetry parameter can be written as
\begin{eqnarray}
\frac{\partial^2}{\partial a^2} \frac{\langle\mathcal{S}_{11}\rangle}{k_BT\langle g\rangle}&=& f(a,\Gamma_{1},\Gamma_{2}) \times g(\Phi,\Delta), \label{second}
\end{eqnarray}
where $f(a,\Gamma_{1},\Gamma_{2})\equiv 4+3\Gamma_{1}(a-1)-3\Gamma_{2}(a+1)$ and $g(\Phi,\Delta)$ is a function of $\Phi$ and of $\Delta$. We observe that the sign of the second derivative is fixed by the sign of $f$. We separate  the diagram generated by $\Gamma_{1} \times \Gamma_{2}$  into three distinct regions according to the sign $(+)$ and $(-)$ as is exhibited in figure \ref{second2}. The $(+)$ and $(-)$ regions determine, respectively, upward or downward concavity, whereas $(+/-)$ determines a change of concavity in the sign of the noise power in $a \in [-1,1]$.  Note that these regions are separated be the straight lines $\Gamma_{1}=2/3 $ and $\Gamma_{2}=2/3$ in the diagram.  The particular case of ideal, maximum tunneling rate, case is a vertex of the diagram situated in the $(-)$ region, whereas the opaque, zero tunneling, limit is close to the vertex in the neighborhood of the $(+)$ region in the diagram.

\begin{figure}
\begin{center}
\includegraphics[width=10.0cm,height=7.0cm]{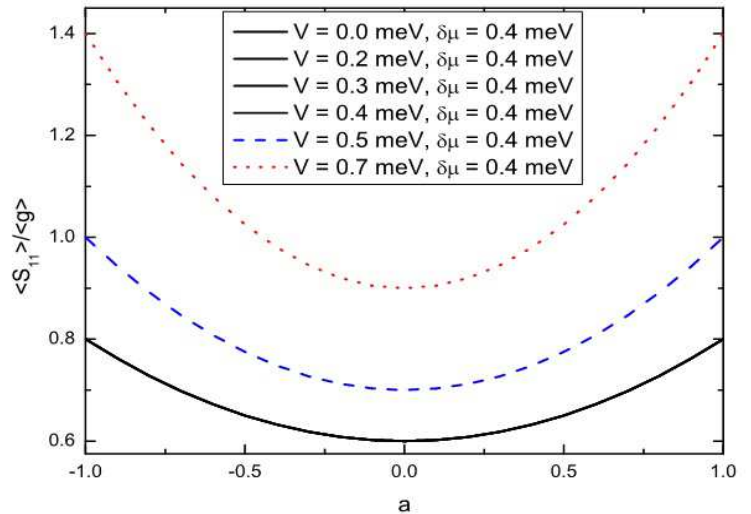}
\end{center}
\caption{We show the behavior of the shot noise in the case of spin accumulation as a function of  $a=\left(G_1-G_2\right)/G_T$, in a QD with non-ideal contacts, equation (\ref{shotnoisespin}). For any value such that $\delta\mu \geq eV$, the result will always be the same for the shot noise which is a direct consequence of the spin accumulation in the reservoirs.} \label{shotnoise}
\end{figure}

Finally, we consider the limit $ eV, \delta \mu \gg k_BT $ in (\ref{limiteopaco}). This limit allows us to get the shot-noise given by the following expression
\begin{eqnarray}
\frac{\langle\mathcal{S}_{11}\rangle}{\langle g\rangle}&=&\frac{\left(1+a^2\right)}{2}\left(|eV+\delta\mu|+|eV-\delta\mu|\right) \nonumber \\
&& +\frac{\left(1-a^2\right)}{2}|\delta\mu|; \label{shotnoisespin}\\
\frac{\langle\mathcal{S}_{11}\rangle}{\langle g\rangle}&=&2\frac{\left(1+a^2\right)}{2}|eV|,\quad eV\gg\delta\mu \nonumber\\
\frac{\langle\mathcal{S}_{11}\rangle}{\langle g\rangle}&=&2\frac{\left(3+a^2\right)}{4}|\delta\mu|,\quad \delta\mu\gg eV\nonumber
\end{eqnarray}
One of the principal result of this paper is the following: In systems with accumulation of the spin, the Fano factor, $ F = \left (1+ a ^ 2 \right) / 2 $, measured experimentally in Ref.~ [\onlinecite{gustavson1}] without taking into account the spin accumulation, presents a correction given by $ \left (1-a^2 \right) / 4 $. Thus the Fano factor changes to $ F = \left (3 + a^ 2 \right) / 4 $ in the limit $ \delta \mu \gg $ eV with $ F = \langle \mathcal {S} _ {11} \rangle/2eI$. Figure \ref{shotnoise} shows the shot-noise, equation (\ref{shotnoisespin}), as a function of the parameter $ a $. For any value where $ \delta \mu \geq $ eV the result for the shot-noise will always be the same, once again indicating a saturation of the spin accumulation in the noise power. When $ eV \geq \delta \mu $ the shot-noise power approaches the result without spin accumulation.\\

\section{Summary and Conclusions}

In this paper we have analyzed the effect of tunneling and reflection at the gates of open quantum dots on the spin accumulation in electronic reservoirs. We analyzed separately the spin-up and spin-down Fermi distributions, and studied the average current-current correlation function using the Landauer-B\"{u}ttiker formalism. More specifically we investigated noise power in the presence of reflection at the voltage gates of the QD using the Poisson kernel as the scattering matrix distribution. We have obtained general equation for the study of the multi-terminal case with spin accumulation at the thermal crossover, and gave details for the two-terminal case. The dominant term in the semi-classical expansion of the noise power which is valid for all Universal Classes of Random Matrix Ensembles, and in all limits, was shown to be greatly affected by spin accumulation in the reservoirs.

We found that important modification in the behavior of the average noise ensues when tunneling rates are taken into account, especially close to the opaque limit. In particular,by introducing the asymmetry parameter, we have shown the symmetrization of the noise power in the opaque limit of the thermal crossover. We performed a complete analysis of the rather surprising change of the concavity of the trace of the noise as a function of the asymmetry parameter, and have shown that only the opaque limit is totally symmetrical with well defined concavity. We have also exhibited results showing the effect of the tunneling rate on the saturation of the spin accumulation, potentially of experimental value as it shows the effects of the induced potentials due to the spin accumulation.

In Ref.~[\onlinecite{gossard}], it was shown that fine adjustment of the voltage gates can alter the orbital configuration of the QD, restauring the tunneling between resonant levels of excited spin states in the presence of a magnetic field. An asymmetry parameter was used in Ref.~[\onlinecite{gustavson1}] to obtain the noise in the presence of finite tunneling rates. Typical values used in that reference were in the range 1000 Hz - 10000 Hz, generating clear noise signal as a function of the asymmetry parameter. We performed an analysis of the correction to the shot-noise power and the Fano factor, resulting from the spin accumulation, in terms of reflections at the gates and the asymmetry parameter. Our findings may facilitate the experimental study of spin accumulation in reservoirs of mesoscopic systems in general. Other recent studies including electron-electron interaction or capacitance can be investigated considering barriers and spin accumulations [\onlinecite{more1,more2}].

\appendix*

\begin{acknowledgments}
This work was partially supported by CNPq, FAPESP, FACEPE, and INCT-IQ  (Brazilian Agencies).
\end{acknowledgments}

\end{document}